# Distributed Detection of Sparse Stochastic Signals via Fusion of 1-bit Local Likelihood Ratios

Chengxi Li, You He, Xueqian Wang, Gang Li, *Senior Member, IEEE*, Pramod K. Varshney, *Life Fellow, IEEE*

*Abstract*—In this letter, we consider the detection of sparse stochastic signals with sensor networks (SNs), where the fusion center (FC) collects 1-bit data from the local sensors and then performs global detection. For this problem, a newly developed 1-bit locally most powerful test (LMPT) detector requires $3.3Q$ sensors to asymptotically achieve the same detection performance as the centralized LMPT (cLMPT) detector with $Q$ sensors. This 1-bit LMPT detector is based on 1-bit quantized observations without any additional processing at the local sensors. However, direct quantization of observations is not the most efficient processing strategy at the sensors since it incurs unnecessary information loss. In this letter, we propose an improved-1-bit LMPT (Im-1-bit LMPT) detector that fuses local 1-bit quantized likelihood ratios (LRs) instead of directly quantized local observations. In addition, we design the quantization thresholds at the local sensors to ensure asymptotically optimal detection performance of the proposed detector. We show that, with the designed quantization thresholds, the proposed Im-1-bit LMPT detector for the detection of sparse signals requires less number of sensor nodes to compensate for the performance loss caused by 1-bit quantization. Simulation results corroborate our theoretical analysis and show the superiority of the proposed Im-1-bit LMPT detector.

## I. Introduction

UNDER the framework of compressed sensing (CS), a sparse signal can be reconstructed perfectly with only a small number of measurements [1]-[3]. In addition to signal recovery, the problem of detecting sparse signals is also important in many applications, such as radar surveillance [4] and cognitive radios [5]. Existing studies have demonstrated that a reliable decision about the absence or presence of sparse signals can be made without complete signal recovery [6],[7].

Advances in CS have resulted in new frameworks for the problem of sparse signal detection with sensor networks (SNs), since instead of high-dimensional data, compressed measurements can be transmitted within the network to save resource usage [8]-[14]. Further, when the bandwidth in the network is extremely limited [39]-[41], only 1-bit data are transmitted to the fusion center (FC) which makes the global decision. In [36], a detector based on generalized likelihood ratio test (GLRT) is proposed for distributed detection of an unknown scalar deterministic signal with 1-bit data. For the detection of sparse signals, compressed measurements at the local sensors can be quantized into 1-bit data and then sent to the FC [15],[16]. In [15], a 1-bit detector based on GLRT is proposed for distributed detection of sparse deterministic signals. In [16], Bernoulli-Gaussian distribution is used to model the sparse signals and a 1-bit detector based on locally most powerful test (LMPT) is proposed for the distributed detection of sparse stochastic signals. The 1-bit LMPT detector in [16] does not require signal recovery, which is more computationally efficient than the 1-bit GLRT detector in [15] based on complete reconstruction of sparse signals. It is worth mentioning that, by fusing directly quantized measurements without any processing, the 1-bit LMPT detector [16] with $3.3Q$ sensors can attain the same detection performance as the centralized LMPT (cLMPT) detector [12] that fuses analog measurements from $Q$ sensors.

In addition to detectors based on directly quantized data like [16], other detectors based on quantized likelihood ratios (LRs) have been studied in the existing literature [17]-[22] for the classical problems of distributed detection, which demonstrate that information loss induced by local quantization can be minimized by optimal quantization of LRs. However, the local quantization of LRs in the context of distributed detection of sparse signals has not been investigated in the literature.

In this letter, we first propose an improved-1-bit LMPT (Im-1-bit LMPT) detector for distributed detection of sparse signals, which is a modified version of the 1-bit LMPT detector proposed in [16]. The key modification[1] is that, at the local sensors, LRs are quantized to generate the 1-bit data, instead of directly quantizing the analog observations as in [16]. We show that quantization of LRs is optimal in the asymptotical regime for the distributed detection problem considered in this letter. It is worth emphasizing that different processing methods at the local sensors lead to significantly different detectors at the FC, even though LMPT strategy is adopted at the FC to make the global decision for both detectors. Besides, we design the local quantization thresholds to guarantee its asymptotically optimal detection performance. With the designed 1-bit quantization thresholds at the local sensors, the proposed Im-1-bit LMPT detector with $1.53Q$ sensors can achieve the same detection

This work was supported in part by National Natural Science Foundation of China under Grants 61790551 and 61790554, and in part by National Science Foundation of USA under Grant No. ENG 60064237. Corresponding author: You He. Email: heyou@mail.tsinghua.edu.cn.

C. Li, X. Wang and G. Li are with the Department of Electronic Engineering, Tsinghua University, Beijing, 100084, China.

Y. He is with the Institute of Information Fusion, Naval Aeronautical University, Yantai 264001, China, also with the Department of Electronic Engineering, Tsinghua University, Beijing, 100084, China.

P. K. Varshney is with the Department of Electrical Engineering and Computer Science, Syracuse University, Syracuse, NY 13244, USA.

---

[1]Note that the extension from direct quantization of measurements to LR quantization can also be applied to other 1-bit detectors (e.g., 1-bit GLRT detector in [15]). In this letter, we specifically consider 1-bit LMPT detector in [16] due to its computational simplicity.

performance as the cLMPT detector with $Q$ sensors, which outperforms the 1-bit LMPT detector in [16]. Simulations show the superiority of the proposed Im-1-bit LMPT detector.

## II. SIGNAL MODEL AND BACKGROUND

### A. Signal Model

Suppose there are $Q$ local sensors and an FC in the sensor network (SN). The problem of distributed detection of sparse stochastic signals can be formulated as the following binary hypothesis testing problem [12], [15]:

$$\begin{cases} H_0: y_q = w_q, & q = 1, 2, ..., Q, \\ H_1: y_q = \mathbf{h}_q^T \mathbf{s}_q + w_q, & q = 1, 2, ..., Q, \end{cases} \quad (1)$$

where the subscript $q$ is the sensor index, $w_q$ is the zero-mean Gaussian noise with known variance[2] $\sigma_w^2$, $\mathbf{s}_q \in \mathbb{R}^{N \times 1}$ denotes the sparse vector containing only a few nonzero entries, $\mathbf{h}_q \in \mathbb{R}^{N \times 1}$ is the known gain vector, $y_q$ is the compressed observation, and $(.)^T$ represents the transpose operation. We assume that $\{w_1, w_2, \cdots, w_Q\}$ are independently and identically distributed (i.i.d.) random variables (RVs). Throughout this letter, it is assumed that all the quantities are real-valued.

In what follows, the Bernoulli-Gaussian distribution imposed on the sparse signals $\{\mathbf{s}_q, q = 1, 2, ..., Q\}$ is introduced, which has been widely used in the literature to model sparsity in various applications [7],[12],[16],[37], such as the sparse spectrum occupancy problem in the frequency and time domains of cognitive radio networks [37]. First, an $N \times 1$ vector $\mathbf{u}$ is used to define the joint sparsity pattern of $\{\mathbf{s}_q, q = 1, ..., Q\}$, where $u_n = 1$ indicates $\{s_{q,n}, q = 1, ..., Q\}$ are nonzero values and $u_n = 0$ otherwise, $n = 1, 2, ..., N$. We assume that the entries in $\mathbf{u}$ are i.i.d. Bernoulli RVs, where

$$u_n = \begin{cases} 1, & \text{with probability } p, \\ 0, & \text{with probability } 1 - p, \end{cases} \quad \forall n. \quad (2)$$

Note that the parameter $p$ is defined as the sparsity level of the sparse signals [7]. The nonzero entries in $\{\mathbf{s}_q, q = 1, 2, ..., Q\}$ are assumed to be i.i.d. Gaussian RVs with probability density function (PDF) $\mathcal{N}(0, \sigma_0^2)$. Then, the PDF of the entries in $\{\mathbf{s}_q, q = 1, 2, ..., Q\}$ can be expressed as

$$s_{q,n} \sim p\mathcal{N}(0, \sigma_0^2) + (1-p)\delta(s_{q,n}), n = 1, 2, ..., N, \quad (3)$$

where $\delta(.)$ denotes the Dirac delta function. Here, the signal parameters $\{p, \sigma_0^2\}$ are assumed to be unknown.

From (1) and (3), the PDFs of $y_q$ are

$$y_q | H_0 \sim \mathcal{N}(0, \sigma_w^2), \text{ and } y_q | H_1 \overset{a}{\sim} \mathcal{N}(0, p\sigma_0^2 \|\mathbf{h}_q\|_2^2 + \sigma_w^2), \quad (4)$$

respectively, where 'a' represents 'asymptotic', $\forall q$. The asymptotic PDF of $y_q$ under $H_1$ in (4) is derived based on Lyapounov-Central Limit Theorem and the assumption that $N$ is very large [7].

### B. Related Background

Due to the extremely limited bandwidth in the SN, each sensor is allowed to transmit 1-bit data to the FC. Upon receiving the 1-bit data from all the sensors, the FC makes a global decision about the presence or absence of the sparse signals. In the 1-bit LMPT detector in [16], the FC fuses the directly quantized 1-bit observations received from all the local sensors based on the LMPT strategy. The direct quantization method at the local sensors adopted by [16] is

$$z_q = \text{sign}(y_q - \zeta_q), \quad (5)$$

where $z_q$ is the quantized local 1-bit measurement, $\zeta_q$ denotes the quantization threshold, and the output of sign(.) equals 1 if the given value is positive and 0 otherwise, $\forall q$.

## III. THE PROPOSED IM-1-BIT LMPT DETECTOR AND PERFORMANCE ANALYSIS

In this section, we propose the Im-1-bit LMPT detector that fuses 1-bit local LRs instead of directly quantized measurements for distributed detection of sparse signals. In addition, the quantization thresholds at the local sensors are designed to guarantee asymptotically optimal detection performance of the proposed detector.

### A. Quantization of LRs at the Local Sensors

From (4), the LR at the $q$-th sensor can be written as

$$\text{LR}_q(y_q; p) = P(y_q | H_1; p) / P(y_q | H_0) = c_{0,q} \exp\{c_{1,q} y_q^2\}, \forall q, \quad (6)$$

where $P(y_q | H_0)$ and $P(y_q | H_1; p)$ denote the PDFs of $y_q$ under $H_0$ and $H_1$, respectively, and

$$c_{0,q} = \sqrt{\sigma_w^2 / (p\sigma_0^2 \|\mathbf{h}_q\|_2^2 + \sigma_w^2)}, \quad (7)$$

$$c_{1,q} = (p\sigma_0^2 \|\mathbf{h}_q\|_2^2) / [2\sigma_w^2 (p\sigma_0^2 \|\mathbf{h}_q\|_2^2 + \sigma_w^2)]. \quad (8)$$

The 1-bit quantized $\text{LR}_q(y_q; p)$ can be expressed as

$$b_q = \begin{cases} 1, & \text{if } \text{LR}_q(y_q; p) \geq \lambda_q, \\ 0, & \text{if } \text{LR}_q(y_q; p) < \lambda_q, \end{cases} \quad (9)$$

where $b_q$ is the quantized value of the LR that is sent to the FC and $\lambda_q$ denotes the quantization threshold at the $q$-th sensor, $\forall q$. It is observed from (6) that $\text{LR}_q(y_q; p)$ is a monotonically increasing function with respect to $y_q^2$ or $|y_q|$. Therefore, the quantization rule in (9) can be rewritten as:

$$b_q = \begin{cases} 1, & \text{if } |y_q| \geq \tau_q, \\ 0, & \text{if } |y_q| < \tau_q, \end{cases} \quad (10)$$

---

[2] It can be easily measured in the absence of the signals.

where[3] $\tau_q = \sqrt{(1/c_{1,q})\ln(\lambda_q/c_{0,q})} > 0$, $\forall q$. Hereinafter, we will refer to $\tau_q$ as the quantization threshold instead of $\lambda_q$ for convenience. As shown in (10), the quantization of the LRs does not incur any additional computational burden at the local sensors than direct quantization in (5).

Based on the quantization rule in (10) and the distribution of $y_q$ in (4), we can easily derive the probability mass functions (PMFs) of $b_q$ under $H_0$ and $H_1$ as

$$P(b_q|H_0) = [2\Phi(\tau_q/\sigma_w)]^{b_q}[1 - 2\Phi(\tau_q/\sigma_w)]^{1-b_q}, \quad (11)$$

and

$$P(b_q|H_1;p) = \{2\Phi[f_q(\tau_q,p)]\}^{b_q}\{1 - 2\Phi[f_q(\tau_q,p)]\}^{1-b_q}, (12)$$

for $q=1,2,\ldots,Q$, where $\Phi(x) = \frac{1}{\sqrt{2\pi}}\int_x^{+\infty}\exp(-t^2/2)dt$ and $f_q(x,t) = x/\sqrt{t\sigma_0^2\|\mathbf{h}_q\|_2^2 + \sigma_w^2}$.

### B. Formulation of the Proposed Im-1-Bit LMPT Detector

In this subsection, we introduce the proposed Im-1-bit LMPT detector based on the 1-bit quantized LRs $\mathbf{b} = \{b_1, b_2, \ldots, b_Q\}$. Since the unknown sparsity level $p$ is close to zero and positive under $H_1$ and $p$ equals 0 under $H_0$, the problem of detection of sparse signals is equivalent to the close and one-sided hypothesis testing problem as:

$$\begin{cases} H_0: & p = 0, \\ H_1: & p \to 0^+. \end{cases} \quad (13)$$

Similar to [16], the LMPT strategy is considered here, because it has been proven to be a powerful tool for close and one-sided hypothesis testing problems [12],[23]-[29]. Based on the LMPT strategy, the proposed Im-1-bit LMPT detector is

$$T_{\text{Im-1-bit}} = \left\{[\partial \ln P(\mathbf{b}|H_1;p)/\partial p]/\sqrt{\text{FI}_{\text{Im-1-bit}}(p)}\right\}\bigg|_{p=0} \overset{H_1}{\underset{H_0}{\gtrless}} \eta, \quad (14)$$

where $\eta$ is the global decision threshold at the FC. In (14), $\partial \ln P(\mathbf{b}|H_1;p)/\partial p$ can be derived as

$$\partial \ln P(\mathbf{b}|H_1;p)/\partial p = \partial \sum_{q=1}^Q \ln P(b_q|H_1;p)/\partial p$$

$$\propto \sum_{q=1}^Q b_q \frac{\tau_q\left(p\sigma_0^2\|\mathbf{h}_q\|_2^2 + \sigma_w^2\right)^{-3/2}\|\mathbf{h}_q\|_2^2 \exp\left\{-[f_q(\tau_q,p)]^2/2\right\}}{\{0.5 - \Phi[f_q(\tau_q,p)]\}\Phi[f_q(\tau_q,p)]}, \quad (15)$$

and $\text{FI}_{\text{Im-1-bit}}(p)$ denotes the Fisher information:

$$\text{FI}_{\text{Im-1-bit}}(p) = \mathbb{E}\left[\left(\partial \ln P(\mathbf{b}|H_1;p)/\partial p\right)^2\right]$$

$$= \sum_{q=1}^Q \frac{g[f_q(\tau_q,p)]\|\mathbf{h}_q\|_2^4\sigma_0^4 / \left[4\left(p\sigma_0^2\|\mathbf{h}_q\|_2^2 + \sigma_w^2\right)^2\right]}{\{0.5 - \Phi[f_q(\tau_q,p)]\}\Phi[f_q(\tau_q,p)]}, \quad (16)$$

---

[3] If the threshold is not positive, i.e., $\tau_q \leq 0$, the quantization result $b_q$ will always be 1, which indicates that the 1-bit data cannot provide any information about the signal to be detected.

where $g(x) = (x^2/2\pi)\exp(-x^2)$. Note that, after substituting $p=0$ into (15) and (16), the Im-1-bit LMPT detector can be expressed as

$$T_{\text{Im-1-bit}} \propto \sum_{q=1}^Q \frac{\tau_q\|\mathbf{h}_q\|_2^2 \exp[-\tau_q^2/(2\sigma_w^2)]}{[1/2 - \Phi(\tau_q/\sigma_w)]\Phi(\tau_q/\sigma_w)} b_q \overset{H_1}{\underset{H_0}{\gtrless}} \eta', \quad (17)$$

where $\eta'$ is the modified global decision threshold at the FC.

### C. Design of Quantization Thresholds at the Local Sensors

According to [29], the test statistic $T_{\text{Im-1-bit}}$ in (14) asymptotically follows

$$T_{\text{Im-1-bit}} \overset{a}{\sim} \begin{cases} \mathcal{N}(0,1) & \text{under } H_0, \\ \mathcal{N}(\mu_{\text{Im-1-bit}},1) & \text{under } H_1, \end{cases} \quad (18)$$

for a large value of $Q$, where $\mu_{\text{Im-1-bit}}$ is expressed as

$$\mu_{\text{Im-1-bit}} = p\sqrt{\text{FI}_{\text{Im-1-bit}}(p)} \approx p\sqrt{\text{FI}_{\text{Im-1-bit}}(0)}. \quad (19)$$

From (14) and (18), the global decision threshold $\eta$ at the FC can be easily derived by $\eta = \Phi^{-1}(P_{fa})$, where $P_{fa}$ is the probability of false alarm. It can be noted that the computation of the test threshold at the FC does not require the knowledge of the unknown parameters, i.e., $p$ and $\sigma_0^2$. As indicated by (18) and (19), the Im-1-bit LMPT detector asymptotically achieves better detection performance with increasing values of $\text{FI}_{\text{Im-1-bit}}(0)$. Therefore, the local quantization thresholds $\{\tau_1, \tau_2, \ldots, \tau_Q\}$ can be designed by solving the following optimization problem:

$$\max_{\tau_1,\tau_2,\ldots,\tau_Q} \text{FI}_{\text{Im-1-bit}}(0), \text{ s.t. } \tau_1 > 0, \tau_2 > 0, \ldots, \tau_Q > 0, \quad (20)$$

to achieve the asymptotically optimal detection performance of the proposed detector. It is observed from (16) that $\text{FI}_{\text{Im-1-bit}}(0) = \sum_{q=1}^Q \Psi_q \|\mathbf{h}_q\|_2^4 \sigma_0^4 / (4\sigma_w^4)$, where

$$\Psi_q = g(\tau_q/\sigma_w)/\{[1/2 - \Phi(\tau_q/\sigma_w)]\Phi(\tau_q/\sigma_w)\}, \quad (21)$$

where $g(x)$ has been defined after (16). Thus, the problem in (20) can be decoupled into $Q$ independent optimization problems as

$$\max_{\tau_q} \Psi_q, \quad \text{s.t. } \tau_q > 0, \forall q. \quad (22)$$

Since the objective functions in (22) are not concave, we resort to the particle swarm optimization (PSO) algorithm to find solutions to the $Q$ optimization problems in (22), which does not require convexity or concavity property of the objective function [30]-[32]. Besides, PSO has low computational complexity [32] and the global convergence of PSO can be guaranteed by several approaches such as the trust-region methods [33],[34]. To avoid local optimum solutions, we run the PSO multiple times starting with different initializations. The results of PSO converge to the following point:

$$\hat{\tau}_q \approx 1.482\sigma_w, \forall q, \quad (23)$$

which is valid for all the local sensor nodes. Substituting (23) into (16), the maximum value of the mean value $\mu_{\text{Im-1-bit}}$ in (19) can be derived as

$$\mu_{\text{Im-1-bit}}^{\max} \approx p\sqrt{0.3261\sum_{q=1}^{Q}\|\mathbf{h}_q\|_2^4 \sigma_0^4/\sigma_w^4}. \quad (24)$$

*Remark:* Based on the definition of Fisher information in (16), $\text{FI}_{\text{Im-1-bit}}(0)$ can be rewritten as

$$\text{FI}_{\text{Im-1-bit}}(0) = \sum_{q=1}^{Q}\left(\frac{\left[\partial P(b_q=1|H_1;p)/\partial p\right]^2\big|_{p=0}}{\left[P(b_q=1|H_0)\right]\left[1-P(b_q=1|H_0)\right]}\right)$$

$$\stackrel{\langle 1 \rangle}{\approx} \sum_{q=1}^{Q}\frac{\left[P(b_q=1|H_1;p)-P(b_q=1|H_0)\right]^2}{\left[P(b_q=1|H_0)\right]\left[1-P(b_q=1|H_0)\right]p^2}, \quad (25)$$

where $\langle 1 \rangle$ is derived by the first order Taylor's series expansion [38]. It is known that for a fixed probability of false alarm, the maximum probability of detection can be achieved with LR tests under the Neyman-Pearson criterion [29]. Since the 1-bit quantization at local sensors can be regarded as a local decision test, we can deduce that for a fixed $P(b_q=1|H_0)$, the maximum value of $P(b_q=1|H_1;p)$ is attained when $b_q$ is the quantized value of LRs. Therefore, from (25), the maximum value of $\text{FI}_{\text{Im-1-bit}}(0)$ and the optimal detection performance at the FC can be obtained by quantization of LRs.

### D. Comparison with the Existing LMPT Detectors

The cLMPT detector that fuses analog measurements and the 1-bit LMPT detector have been studied in [12] and [16]:

$$T_c \stackrel{a}{\sim} \begin{cases} \mathcal{N}(0,1) & \text{under } H_0, \\ \mathcal{N}(\mu_c,1) & \text{under } H_1, \end{cases} \quad (26)$$

$$T_{\text{1-bit}} \stackrel{a}{\sim} \begin{cases} \mathcal{N}(0,1) & \text{under } H_0, \\ \mathcal{N}(\mu_{\text{1-bit}},1) & \text{under } H_1, \end{cases} \quad (27)$$

where

$$\mu_c \approx p\sqrt{0.5\sum_{q=1}^{Q}\|\mathbf{h}_q\|_2^4 \sigma_0^4/\sigma_w^4}, \quad (28)$$

and the maximum value of $\mu_{\text{1-bit}}$ with the near optimal local quantization thresholds is [16]

$$\mu_{\text{1-bit}}^{\max} \approx p\sqrt{0.1521\sum_{q=1}^{Q}\|\mathbf{h}_q\|_2^4 \sigma_0^4/\sigma_w^4}. \quad (29)$$

From (28) and (29), in the homogenous scenario where $\|\mathbf{h}_q\|_2^2$ across different sensor nodes are all equal to each other, it has been derived in [16] that

$$\mu_c = \mu_{\text{1-bit}}^{\max} \Rightarrow Q_{\text{1-bit}} \approx 3.3 Q_c, \quad (30)$$

where $Q_{\text{1-bit}}$ and $Q_c$ are the number of sensors required by the 1-bit LMPT detector and the cLMPT detector to asymptotically achieve the same detection performance, respectively. From (24) and (28), when the cLMPT detector and the proposed Im-1-bit LMPT detector have the same asymptotical detection performance in the homogenous scenario, we can derive that

$$\mu_c = \mu_{\text{Im-1-bit}}^{\max} \Rightarrow Q_{\text{Im-1-bit}} = \frac{0.5}{0.3261}Q_c \approx 1.53 Q_c, \quad (31)$$

where $Q_{\text{Im-1-bit}}$ is the number of sensors required by the Im-1-bit LMPT detector. Therefore, from (30) and (31), one can observe that the proposed Im-1-bit LMPT detector requires fewer sensors to compensate for the detection performance loss induced by quantization compared with the 1-bit LMPT detector [16].

## IV. SIMULATION RESULTS

In this section, simulation results are provided to demonstrate the performance of the proposed Im-1-bit LMPT detector. For all the simulations, the elements in $\{\mathbf{h}_q, q=1,2,...,Q\}$ are sampled from i.i.d. standard normal distributions and then normalized to satisfy $\|\mathbf{h}_q\|_2^2 = 1, \forall q$.

In Fig. 1, we plot the receiver operating characteristic (ROC) curves for the Im-1-bit LMPT detector and the 1-bit LMPT detector through $10^4$ Monte Carlo trials, where $Q=300, N=1000, \sigma_w^2=1$, and $p=0.05$. It can be observed that the detection performance of the Im-1-bit LMPT detector is better than that of the 1-bit LMPT detector, which results from LR quantization at the local sensor nodes.

In Fig. 2, we plot the ROC curves of the Im-1-bit LMPT detector and the cLMPT detector for different number of sensors and different sparsity levels $p$, where $\sigma_w^2=1$, $\sigma_0^2=8$, and $N=1000$. From Fig. 2, the Im-1-bit LMPT detector with 153 sensors approximately achieves the same detection performance as the cLMPT detector with 100 sensors, which is consistent with (31) in Section III and shows the superiority of the proposed detector compared with the 1-bit LMPT detector [16].

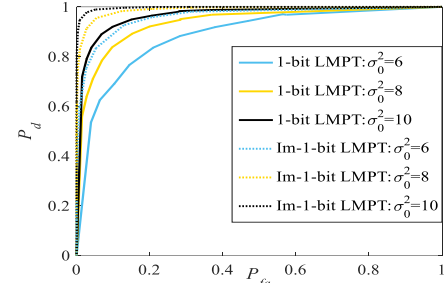

Fig. 1. ROC curves of the Im-1-bit LMPT detector and the 1-bit LMPT detector, where $\sigma_w^2=1$, $Q=300, N=1000$ and $p=0.05$.

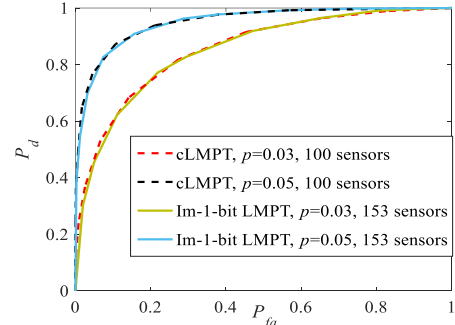

Fig. 2. ROC curves of the Im-1-bit LMPT detector and the cLMPT detector, where $\sigma_w^2=1, \sigma_0^2=8$, and $N=1000$.

## V. CONCLUSION

The problem of distributed detection of sparse stochastic signals with 1-bit data was considered in this letter. We proposed the Im-1-bit LMPT detector as an improved version of the existing 1-bit LMPT detector by fusing the local quantized

LRs instead of the directly quantized observations and derived the asymptotically optimal local quantization thresholds. It is shown theoretically and numerically that the Im-1-bit LMPT detector with $1.53Q$ sensors can asymptotically achieve the same detection performance as the cLMPT detector with $Q$ sensors, which outperforms the original 1-bit LMPT detector. It is worth noting that the Im-1-bit LMPT detector proposed in this letter can be easily extended to the case where multiple-level data are transmitted by the local sensors. Further avenues of study will include the Im-1-bit LMPT detector dealing with sparse signals that are correlated across different nodes [35].